\documentclass[aps,prb,reprint,english,a4paper,floatfix,showpacs,10pt,superscriptaddress]{revtex4-1}
\usepackage[T1]{fontenc}
\usepackage{ae,aecompl}
\usepackage[english]{babel}
\usepackage{mathtools}
\usepackage{url}
\usepackage{psfrag,color,pstricks,pst-grad}
\usepackage[dvips]{graphicx}
\usepackage{amsmath,amsfonts,amssymb,amsbsy}  
\usepackage{hyperref}
\usepackage[normalem]{ulem}

\newcommand{\beq}{\begin{equation}}
\newcommand{\eeq}{\end{equation}}

\begin{document}

\title{Nonequilibrium phonon mean free paths in anharmonic chains}
\author{K. S\"a\"askilahti}
\email{kimmo.saaskilahti@aalto.fi}
\author{J. Oksanen}
\affiliation{Department of Biomedical Engineering and Computational Science, Aalto University, FI-00076 AALTO, FINLAND}
\author{S. Volz}
\affiliation{Ecole Centrale Paris, Grande Voie des Vignes, 92295 Ch\^atenay-Malabry, France}
\affiliation{CNRS, UPR 288 Laboratoire d'Energ\'etique Mol\'eculaire et Macroscopique, Combustion (EM2C), Grande Voie des Vignes, 92295 Ch\^atenay-Malabry, France}
\author{J. Tulkki}
\affiliation{Department of Biomedical Engineering and Computational Science, Aalto University, FI-00076 AALTO, FINLAND}
\date{\today}

\begin{abstract}
 Harnessing the power of low-dimensional materials in thermal applications calls for a solid understanding of the anomalous thermal properties of such systems. We analyze thermal conduction in one-dimensional systems by determining the frequency-dependent phonon mean free paths (MFPs) for an anharmonic chain, delivering insight into the diverging thermal conductivity observed in computer simulations. In our approach, the MFPs are extracted from the length-dependence of the spectral heat current obtained from nonequilibrium molecular dynamics simulations. At low frequencies, the results reveal a power-law dependence of the MFPs on frequency, in agreement with the diverging conductivity and the recently determined equilibrium MFPs. At higher frequencies, however, the nonequilibrium MFPs consistently exceed the equilibrium MFPs, highlighting the differences between the two quantities. Exerting pressure on the chain is shown to suppress the mean free paths and to generate a weaker divergence of MFPs at low frequencies. The results deliver important insight into anomalous thermal conduction in low-dimensional systems and also reveal differences between the MFPs obtained from equilibrium and nonequilibrium simulations. 
\end{abstract}
 \maketitle


\section{Introduction}

Advances in nanotechnologies have facilitated the control of thermal energy \cite{cahill03,cahill14}, enabling, e.g., designing more efficient thermoelectric devices \cite{vineis10}, reducing heating in electronic devices \cite{pop10}, and processing information using phonons \cite{li12_rmp}. It has therefore become vital to properly understand thermal conduction not only in bulk materials but also in low-dimensional systems such as nanowires \cite{chang06_prl,chang08} and membranes \cite{xu14}. While it is well understood that Fourier's classical law of conduction breaks down in small systems due to partially ballistic conduction, there are still several open questions related to the so-called anomalous thermal conduction \cite{lepri03,dhar08} in low-dimensional systems, characterized by the divergence of thermal conductivity $\kappa$ as a function of system size $N$ for $N\to \infty$. 

Much of the theoretical work on anomalous thermal conduction has been carried out for the simplest model system available, a one-dimensional anharmonic chain. Nonequilibrium molecular dynamics (NEMD) simulations have shown \cite{lepri97} that the conductivity diverges as $\kappa(N)\propto N^{\alpha}$, where values $\alpha=1/3$ \cite{mai07} and $\alpha=0.4$ \cite{wang11} have been suggested for the Fermi-Pasta-Ulam (FPU) potential \cite{fermi55} with a quartic nonlinearity. While the value $\alpha=1/3$ is supported by hydrodynamics \cite{narayan02,mai06}, the value $\alpha=0.4$ is in agreement with the mode-coupling theory \cite{lepri98,lepri03}. Recently developed nonlinear hydrodynamics theory \cite{mendl13,spohn14} also predicts the value $\alpha=1/3$, with the exception of a chain with a symmetric interaction potential and zero pressure, for which $\alpha=1/2$ is predicted \cite{spohn14,das14}.

Such discrepancies call for complementary and more detailed tools and methods delivering better understanding of heat transfer in low-dimensional systems. Nonequilibrium simulations and hydrodynamics produce powerful predictions but cannot directly provide intuitive understanding of, e.g., the roles played by different vibrational modes in thermal conduction. In a recent work, Liu \textit{et al}. \cite{liu14} provided such mode-level predictions by determining the frequency-dependent mean free paths (MFPs) $\Lambda(\omega)$ in anharmonic chains from the velocity cross-correlation function obtained from equilibrium molecular dynamics, similarly to earlier works on more complex systems (see, e.g., Refs. \cite{thomas10,latour14}). Liu \textit{et al}. found a power-law dependence of MFPs on angular frequency, $\Lambda(\omega)\propto \omega^{-\eta}$, where $\eta \approx 1.70$. Equilibrium \cite{pereverzev03} and non-equilibrium \cite{saaskilahti15} considerations show that such a power-law divergence leads to a thermal conductivity diverging with exponent $\alpha=1-1/\eta\approx 0.41$, in close agreement with the mode-coupling theory. 

In this work, we apply the recently developed alternative method \cite{saaskilahti15} to determine the frequency-dependent phonon MFPs in an anharmonic chain. The method is based on calculating the spectral decomposition of the heat current \cite{saaskilahti14b} for various chain lengths from NEMD simulations and inferring the MFPs from the length-dependence of the heat current at each frequency. The method \textit{directly probes the nonequilibrium MFP corresponding to the decay of the heat flux} instead of the equilibrium relaxation length. These two length scales do not generally agree, as manifested in, e.g., the failure of the relaxation-time approximation in systems with relatively weak Umklapp scattering \cite{lindsay13}. By comparing the nonequilibrium MFPs to the equilibrium MFPs of Liu \textit{et al.} \cite{liu14}, we can obtain a better understanding of the differences between the two quantities. It is also important to note that in contrast to perturbative calculations \cite{lepri98,pereverzev03}, our NEMD simulations inherently account for all orders of phonon-phonon scattering processes and their (complex) effects on the phonon distribution function in nonequilibrium. Although there is no known physical system corresponding to the quartic anharmonic chain studied in this paper, we expect our results and methods to provide useful microscopic insight also to the experimental investigations of anomalous heat conduction in low-dimensional systems such as nanotubes \cite{chang08}, polymers \cite{shen10}, and graphene \cite{xu14}. 

The paper is organized as follows. Section \ref{sec:theory} defines the spectral decomposition of heat current and shows how the nonequilibrium MFPs are determined. Section \ref{sec:results} presents numerical results for the $\beta$-FPU chain, showing the dependence of the spectral current on the system size and the frequency-dependent MFPs. We also investigate the effects of external pressure on the MFPs. We conclude in Sec. \ref{sec:conclusions}.

\section{Theory}
\label{sec:theory}
\subsection{Spectral heat current}
The spectral decomposition of heat current can be found by monitoring the force-velocity correlations between neighboring particles as described in Ref. \cite{saaskilahti14b} and briefly reviewed here. The starting point for the analysis is the average heat current $Q_{i \to i+1}$ flowing between neighboring particles in the chain in thermal nonequilibrium, with the particles labeled by indices $i$ and $i+1$. The microscopic expression for $Q_{i \to i+1}$ is given by the statistical average of the rate of work done by the atoms on each other:
\begin{equation}
 Q_{i \to i+1} =  \frac{1}{2} \langle {F}_{i+1,i}  \left[ {v}_{i+1} + {v}_{i}\right] \rangle.
\end{equation}
Here, one-dimensional dynamics has been assumed so that the particle velocities $v_i$ and $v_{i+1}$ are scalar. The interparticle force ${F}_{i+1,i}=-\partial V_{i,i+1}/\partial x_{i+1}$ is obtained from the interaction potential $V_{i,i+1}$ by differentiating with respect to particle position $x_{i+1}$. Following the derivation of Ref. \cite{saaskilahti14b}, the heat current $Q_{i \to i+1}$ can be decomposed spectrally as
\begin{equation}
 Q_{i \to i+1}  = \int_0^{\infty} \frac{d\omega}{2\pi} q_{i \to i+1}(\omega),
\end{equation}
where $\omega$ is the angular frequency and the spectral heat current $q_{i\to i+1}(\omega)$ is given by the expression
\begin{equation}
 q_{i\to i+1}(\omega) = \textrm{Re} \int_{-\infty}^{\infty} d\tau e^{i\omega \tau} \langle {F}_{i+1,i}(\tau) \left[v_{i+1}(0)+ {v}_i(0) \right]  \rangle. \label{eq:qomega}
\end{equation}
Here $\tau$ is the correlation time between forces and velocities and the correlation function $\langle {F}_{i+1,i}(t_1) \cdot [v_{i+1}(t_2)+{v}_i(t_2)]  \rangle$ only depends on the time-difference $t_1-t_2$ due to the assumed steady state.

For time-invariant systems, one can employ the Fourier transform identity
\begin{equation}
 2\pi \delta(\omega+\omega') q_{i \to i+1}(\omega) = \textrm{Re} \langle \tilde{{F}}_{i+1,i}(\omega) [\tilde{{v}}_{i+1}(\omega')+\tilde{v}_{i}(\omega')]  \rangle \label{eq:q_wiener}
\end{equation}
for the calculation of spectral heat current. Here the (generalized) Fourier transforms are defined for, e.g., the velocities as
\begin{equation}
 \tilde{v}_i(\omega) = \int_{-t_c/2}^{t_c/2} dt e^{i\omega t} v_i(t),
\end{equation}
where the cut-off time $t_c$ is introduced to regularize the integral and the limit $t_c\to \infty$ is implied in Eq. \eqref{eq:q_wiener}. Equation \eqref{eq:q_wiener} allows for more direct evaluation of $q_{i \to i+1}(\omega)$ in terms of the (discrete) Fourier transforms of force and velocity trajectories obtained from NEMD simulations. More details of the practical evaluation of $q_{i \to i+1}(\omega)$ in terms of discrete Fourier transformed trajectories and the required spectral smoothing are given below in Sec. \ref{sec:results} and in Ref. \cite{saaskilahti15}.

\subsection{Mean free paths}

Phonon MFPs are determined from the decay of heat current $q_{i\to i+1}(\omega)$ as a function of chain length $N$, where the $N$ atoms are assumed to be sandwiched between two leads at temperatures $T+\Delta T/2$ and $T-\Delta T/2$. The expected functional form of the length-dependence can be inferred from the solution of the Boltzmann transport equation for a one-dimensional chain \cite{saaskilahti13}, giving (particle indices are suppressed for the simplicity of notation)
\begin{equation}
 q(\omega) = \frac{M(\omega)}{1+\frac{N}{2\Lambda(\omega)}}\Delta T, \label{eq:qomega_N}
\end{equation}
where we have set the Boltzmann constant $k_B=1$. Here $M(\omega)$ does not depend on $N$ and $\Lambda(\omega)$ is the phonon MFP at frequency $\omega$. In the nearly ballistic limit (as in the carbon nanotube at room temperature considered in Ref. \cite{saaskilahti15}), quantity $M(\omega)$ is an integer equal to the number of propagating modes. In the anharmonic FPU chain, however, the anharmonicity both renormalizes phonon modes \cite{li06b} and generates a finite phonon life-time, shifting dispersion to higher frequencies and dissipating phonons also in the leads. Therefore, $M(\omega)$ now lacks the simple interpretation as the number of modes and should simply be taken as the spectral current for $N\to 0^+$.

Since the validity of Eq. \eqref{eq:qomega_N} and the value of $M(\omega)$ are unknown \textit{a priori} for the anharmonic chain, Eq. \eqref{eq:qomega_N} is not directly used to determine $\Lambda(\omega)$ by calculating $q(\omega)$ for a single $N$ and solving for $\Lambda(\omega)$. Instead, the MFPs are found by plotting $[q(\omega)/\Delta T]^{-1}$ versus chain length $N$ for various $N$, looking for a linear dependence and determining $\Lambda(\omega)$ from the inverse slope \cite{savic08_prb,savic08_prl,saaskilahti15}. This procedure suppresses the numerical error and also allows for estimating the validity of Eq. \eqref{eq:qomega_N}.

Below, we compare the non-equilibrium MFPs determined from Eq. \eqref{eq:qomega_N} to the equilibrium MFPs \cite{liu14} determined from the decay of velocity cross-correlations at each frequency $\omega$. Because the MFPs reported in Ref. \cite{liu14} correspond to the decay length of vibrational amplitude and not the square of amplitude as in our definition, we divide the mean free paths of Ref. \cite{liu14} by two to allow for meaningful comparison. This procedure ensures that for a harmonic chain coupled to dissipative Langevin heat baths with coupling constant $\gamma$ (see Sec. \ref{sec:res_setup} below), the equilibrium and non-equilibrium methods deliver the same MFP $\Lambda(\omega)=v(\omega)\gamma^{-1}$, where $v(\omega)=\sqrt{1-\omega^2/4}$ is the mode velocity in dimensionless units. 

It is important to note that because the MFPs $\Lambda(\omega)$ determined from nonequilibrium simulations via Eq. \eqref{eq:qomega_N} directly define the length-dependence of the spectral current, the MFPs can be used to estimate the total thermal conductivity $\kappa=QN/\Delta T$ as \cite{saaskilahti15}
\begin{equation}
 \kappa(N) = N \int_0^{\infty} \frac{d\omega}{2\pi} \frac{M(\omega)}{1+\frac{N}{2\Lambda(\omega)}}. \label{eq:kappaN}
\end{equation}
In the definition of $\kappa$, we have used the ''experimental'' definition and divided $Q$ by $\Delta T/N$ instead of the thermal gradient $dT/dx$, because the relation between the current and the gradient is not well defined for an anomalous conductor \cite{liu14_prl}. Equation \eqref{eq:kappaN} allows for easily understanding the origin of anomalous conductivity: assuming a power-law divergence $\Lambda(\omega) \propto \omega^{-\eta}$ of the MFP at low frequencies, assuming $M(\omega)$ to approach a constant, and taking the asymptotic limit $N \to \infty$ gives \cite{saaskilahti15} $\kappa(N)\propto N^{1-1/\eta}$, showing that the MFP divergence exponent $\eta$ and the thermal conductivity divergence exponent $\alpha$ are related as $\alpha=1-1/\eta$. The same relation between the low-frequency mean free paths and the divergence of thermal conductivity has been also suggested using linear response theory \cite{lepri03}, Peierls equation \cite{pereverzev03}, and by treating the finite length of the system as a source of boundary scattering \cite{mingo05_nanolett}.

Finally, we note that when $\Lambda(\omega)$ is bounded from above, one can directly take the limit $N \to \infty$ in Eq. \eqref{eq:kappaN} to get the size-independent thermal conductivity
\begin{equation}
 \kappa = 2 \int_0^{\infty} \frac{d\omega}{2\pi} M(\omega) \Lambda(\omega).
\end{equation}
This expression can be shown to be equivalent to the classical Boltzmann-Peierls expression \cite{ziman} for thermal conductivity, allowing for interpreting $\Lambda(\omega)$ as the ''transport relaxation length''\cite{sun10}. In low-dimensional materials, where the relaxation time approximation is often inadequate \cite{lindsay13}, calculating the transport relaxation length from quantum-mechanical first-principles calculations generally requires the full solution of the Boltzmann equation \cite{broido05}. In our classical NEMD approach, the transport relaxation lengths are obtained directly via Eq. \eqref{eq:qomega_N}.

\section{Numerical results}
\label{sec:results}
\subsection{System setup}
\label{sec:res_setup}

\begin{figure}
 \begin{center}
  \includegraphics[width=8.6cm]{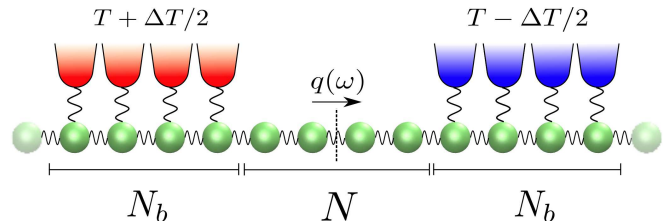}
 \caption{Schematic illustration of the studied system. Nearest-neighbor particles are coupled by anharmonic springs. The left- and right-most particles are fixed, and $N_b$ particles at left and right ends of the system are coupled to Langevin baths at temperatures $T+\Delta T/2$ and $T-\Delta T/2$.}
 \label{fig:geom}
 \end{center}
\end{figure}

The studied system is depicted in Fig. \ref{fig:geom}. An anharmonic chain of $N$ atoms is sandwiched between two thermalized chains of length $N_b$. Nearest-neighbor atoms interact through the $\beta$-FPU potential
\begin{equation}
 V_{ij} = \frac{1}{2} m \omega_0^2 (u_i-u_j)^2 + \frac{\beta}{4} (u_i-u_j)^4,
\end{equation}
where $u_i=x_i-ia_0$ is the displacement of particle $i$ with position $x_i$ from its equilibrium position $ia_0$, $a_0$ is the lattice constant at zero pressure, $m$ is the particle mass, $\omega_0$ is the spring resonance frequency, and $\beta$ is the anharmonicity parameter. For the simplicity of discussion, considerations of asymmetric potentials such as the FPU potential with a cubic non-linearity are left for future work. $N_b$ particles at the left and right of the $N$ atoms in the middle are coupled to stochastic Langevin heat baths at temperatures $T+\Delta T/2$ and $T-\Delta T/2$, respectively. For a particle $i$ coupled to the heat bath, the bath exerts the time-dependent force 
\begin{equation}
 F_i^{\textrm{bath}} = \xi_i - m\gamma v_i,
\end{equation}
where the bath coupling strength $\gamma$ is related to the variance of the stochastic force $\xi_i$ through the classical fluctuation-dissipation relation
\begin{equation}
 \langle \xi_i(t) \xi_i(t') \rangle = 2m \gamma T_i \delta(t-t').
\end{equation}
The left- and right-most atoms in the system of Fig. \ref{fig:geom} are fixed. We have checked that for a chain at zero external pressure, using free boundary conditions delivers the same results as fixed boundary conditions.

We use dimensionless units $t'=\omega_0 t$, $u'=u/\sqrt{m\omega_0^2/\beta}$, $\gamma'=\gamma/\omega_0$, $T'_i=\beta T_i/(m^2\omega_0^4)$ throughout the text, which essentially enables setting the atom mass $m$, oscillator frequency $\omega_0$, and the anharmonicity parameter $\beta$ equal to unity in the equation of motion without loss of generality \cite{saaskilahti12}. The dimensionless temperature $T'$ controls the strength of anharmonicity. Below, the primes denoting the dimensionless units are suppressed. Simulations are performed using the LAMMPS simulation package \cite{plimpton95} with time step $\Delta t=0.005$ and simulation duration of $10^9$ steps, with the exception of the longest chains considered ($N=20000$ and $N=30000$) for which $2\times 10^9$ molecular dynamics steps were used. Spectral heat current $q(\omega)$ is determined from Eq. \eqref{eq:q_wiener} by using the discrete Fourier transformed force and velocity trajectories evaluated at the middle of the chain (dashed line in Fig. \ref{fig:geom}). The sharply fluctuating $q(\omega)$ is spectrally smoothened with a rectangular window of width $\Delta \omega=0.01$ unless noted otherwise. 

We use the bath parameters $\gamma=0.01$ and $N_b=1000$ in the calculations below. The effects of the bath parameters on the spectral currents are discussed in App. \ref{app:A}, where it is shown that choosing $\gamma=0.01$ effectively restricts our analysis to frequencies $\omega \gtrsim 0.01$ because of the overdamping of modes at angular frequencies $\omega \lesssim 0.01$. Because the spectrally smoothened current with smoothing width $\Delta \omega=0.01$ exhibits such overdamping artefacts at frequencies $\omega<0.015$, we only show numerical results for frequencies $\omega > 0.015$. 

\subsection{Spectral current}

\begin{figure}
 \begin{center}
  \includegraphics[width=8.6cm]{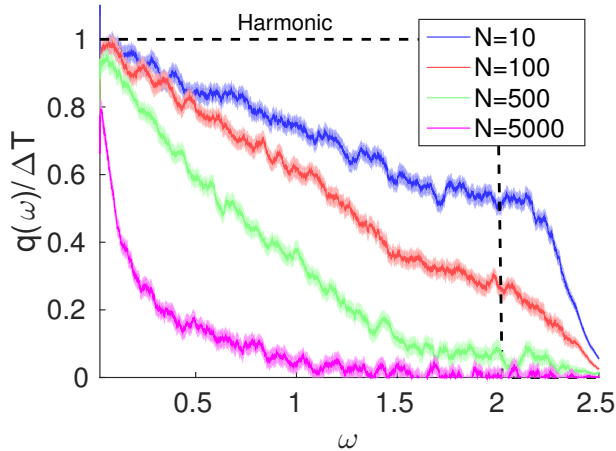}
 \caption{Normalized spectral current $q(\omega)/\Delta T$ for the $\beta$-FPU chain with various chain lengths $N$. Mean temperature is $T=0.2$ and temperature bias $\Delta T=0.1$. Spectral smoothing has been carried out with spectral width $\Delta \omega=0.05$. The shaded regions correspond to the 95\% confidence interval.}
 \label{fig:Toms}
 \end{center}
\end{figure}

Figure \ref{fig:Toms} shows the normalized spectral current $q(\omega)/\Delta T$ at mean bath temperature $T=0.2$ for various chain lengths $N$. The temperature difference of the baths is chosen as $\Delta T=0.1$. We have checked that the currents remain unchanged with $\Delta T=0.05$. For reference, Fig. \ref{fig:Toms} also shows the spectral current for a harmonic chain. In the harmonic case, the normalized current corresponds to the number of propagating modes \cite{saaskilahti15}, equal to unity for $\omega<2$ in dimensionless units. As seen in Fig. \ref{fig:Toms}, the anharmonicity of the $\beta$-FPU potential decreases phonon transmission below unity for $\omega<2$ even for the shortest chain ($N=10$) and also allows for the propagation of heat even above $\omega=2$. The latter effect can be understood based on the effective phonon theory \cite{li06b,li07,liu14}, which predicts that the anharmonicity shifts the dispersion to higher frequencies so that the maximum frequency is $\omega_{\textrm{max}}^{\beta-\textrm{FPU}} = 2\alpha(T)$, where $\alpha(T)\approx 1.175$ for $T=0.2$. In addition to the shift, anharmonicity also broadens the phonon dispersion by introducing a non-zero phonon life-time, reducing the current compared to the harmonic case and allowing for the propagation of heat in the shortest chain even above $\omega>\omega_{\textrm{max}}^{\beta-\textrm{FPU}}$. 

The spectral currents plotted in Fig. \ref{fig:Toms} depend strongly on the chain length $N$ because of the increased phonon damping in long chains. In particular, it can be seen that the damping effect is strongest at high frequencies, suggesting shorter MFP than at low frequencies. In the longest chain considered in Fig. \ref{fig:Toms} ($N=5000$),  low frequencies can be seen to dominate thermal conduction.

\begin{figure}
 \begin{center}
  \includegraphics[width=8.6cm]{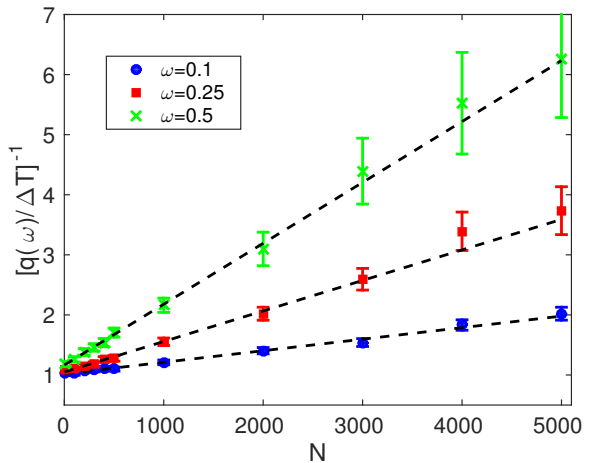}
 \caption{Inverse of the normalized spectral current as a function of chain length $N$ at different frequencies $\omega$. Mean temperature is $T=0.2$ and temperature bias $\Delta T=0.1$. As expected based on the relation $q(\omega)/\Delta T \propto [1+L/2\Lambda(\omega)]^{-1}$, inverse current increases linearly as a function of chain length $N$, allowing for extracting the MFPs from the inverse slopes of the linear fits (dashed lines). The error bars correspond to the estimated 95\% confidence interval.}
 \label{fig:Toms_fit}
 \end{center}
\end{figure}

Equation \eqref{eq:qomega_N} suggests that the inverse of the spectral current is linearly proportional to $N$, so we show in Fig. \ref{fig:Toms_fit} the inverse $[q(\omega)/\Delta T]^{-1}$ of the normalized current as a function of chain length $N$ at different frequencies $\omega$. Linear dependence on $N$ can be seen in the figure, suggesting the validity of Eq. \eqref{eq:qomega_N} in describing the length-dependence of $q(\omega)$. The linear dependence allows for (weighted) linear least squares fitting at each frequency (dashed lines in Fig. \ref{fig:Toms_fit}) and extracting the MFPs $\Lambda(\omega)$ from the inverse slope. 

\subsection{Mean free paths}

\begin{figure}
 \begin{center}
   \includegraphics[width=8.6cm]{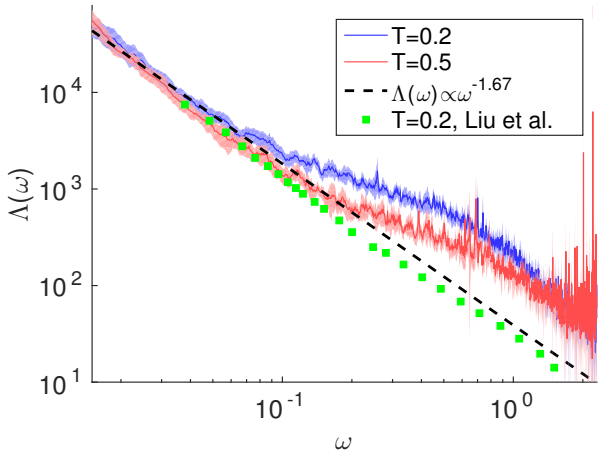}
 \caption{Mean free path $\Lambda(\omega)$ versus frequency at two mean temperatures $T=0.2$ and $T=0.5$. The temperature biases are $\Delta T=0.1$ and $\Delta T=0.25$ for $T=0.2$ and $T=0.5$, respectively. The MFPs were obtained by the linear fitting procedure described in Fig. \ref{fig:Toms_fit}. The fitting was performed by using the current spectra $q(\omega)$ calculated from NEMD simulations for 14 different chain lengths $N$ ranging from $N=10$ to $N=30000$. The MFP data of Liu \textit{et al.}\cite{liu14} have been divided by two to conform to the definition of MFP used in this paper.}
 \label{fig:Lambdas}
 \end{center}
\end{figure}

Phonon MFPs $\Lambda(\omega)$ determined from least squares fitting are plotted in double logarithmic scale in Fig. \ref{fig:Lambdas} for two different temperatures. As expected, increasing the temperature reduces the MFPs due to increasing phonon-phonon scattering. While the effect of temperature on MFPs is relatively small at low frequencies, the effect becomes larger at high frequencies. At low frequencies, MFPs are seen to follow a power-law $\Lambda(\omega) \propto \omega^{-\eta}$ as a function of frequency, and linear least squares fitting for $T=0.2$ and $\omega<0.09$ delivers $\eta\approx 1.67$, which agrees with the mode-coupling theory \cite{lepri98} and predicts thermal conductivity divergence exponent $\alpha=0.40$. Considering that the mode-coupling theory is developed for weak anharmonicity, the close agreement between the theory and our results suggests that either the anharmonicity produced by the temperature $T=0.2$ corresponds to the weak anharmonicity limit or that the scaling exponent derived using the assumption of weak anharmonicity also extends to strong anharmonicities. It is, however, challenging to estimate the statistical error in $\eta$ using standard linear regression methods, because the mean free paths at nearby frequencies are correlated due to the spectral smoothing performed for $q(\omega)$ at each chain length $N$. Reducing the size of the smoothing window $\Delta \omega$ would reduce the correlation, but it would also increase the statistical noise. Directly suppressing the noise by performing longer simulation runs is not, however, feasible with the available computational resources and is therefore left for future work. 

For comparison, Fig. \ref{fig:Lambdas} also includes the MFP data of Liu \textit{et al}. \cite{liu14}, determined for $T=0.2$ from the velocity cross-correlation function in equilibrium. Comparison of the nonequilibrium and equilibrium MFPs shows that the nonequilibrium MFPs are generally larger than the equilibrium ones, with the largest deviations observed at high frequencies. The difference between the two quantities can be understood heuristically by noting that whereas the equilibrium MFPs measure the total relaxation length due to the combined effect of both normal and Umklapp scattering processes, the nonequilibrium MFPs are less affected by the non-resistive normal processes. It has been argued \cite{lindsay13} that equilibrium relaxation lengths correctly describe the decay of heat flux only in systems with sufficiently strong Umklapp scattering, which is the case in, e.g., bulk Si and Ge \cite{ward10}, but not in low-dimensional systems such as graphene \cite{lindsay10c} or carbon nanotubes \cite{lindsay09}. The close agreement of equilibrium and nonequilibrium MFPs at low frequencies seen in Fig. \ref{fig:Lambdas} could therefore be interpreted as the dominance of Umklapp scattering at low frequencies in the $\beta$-FPU chain. All in all, the differences in the equilibrium and nonequilibrium MFPs of Fig. \ref{fig:Lambdas} support our earlier arguments \cite{saaskilahti15} that the MFPs extracted from nonequilibrium heat current generally differ from the equilibrium MFPs.

The longest chains simulated in this work have length $N=30000$. Simulating longer chains would allow for improving the accuracy of MFPs especially at low frequencies, where MFPs exceed 30000 lattice units. Increasing the chain length decreases, however, the heat current flowing in the chain, reducing the statistical accuracy of current spectra. Better statistical accuracy could be obtained by increasing the temperature bias, which may take the system to non-linear regime and is therefore undesirable, or increasing the number of simulation steps beyond $2\times 10^9$ steps. We expect that future speed-ups in computer simulations allow for conveniently simulating chains longer than $N=30000$ for tens of billions of molecular dynamics steps, improving the accuracy of low-frequency nonequilibrium MFPs. 

\subsection{Effect of pressure}

Non-linear hydrodynamics suggests that exerting a non-zero pressure $P$ on the $\beta$-FPU chain takes the system to a different class of thermal conduction ($\alpha=0.33$) than the case of $P=0$ ($\alpha=0.5$). We therefore investigate if this conjencture shows up in the MFPs. 

For the one-dimensional chain, the pressure is simply defined as $P=\langle F_{i+1,i}\rangle$. For the $\beta$-FPU system, zero pressure corresponds to choosing the average nearest-neighbor separation $a=1$, corresponding to the system length $L=N_{tot}$, where $N_{tot}=N+2(N_b+1)$ is the total number of atoms in the chain. Positive and negative pressure can be exerted by reducing or increasing the system length, respectively, which forces the atoms closer ($a<1$) or further away ($a>1$). Because imposing the pressure $-P$ delivers the same spectral current distribution as the pressure $P$ due to the symmetry of the $\beta$-FPU potential, we consider only positive pressures and lattice constants $a \leq 1$ below.

\begin{figure}
 \begin{center}
   \includegraphics[width=8.6cm]{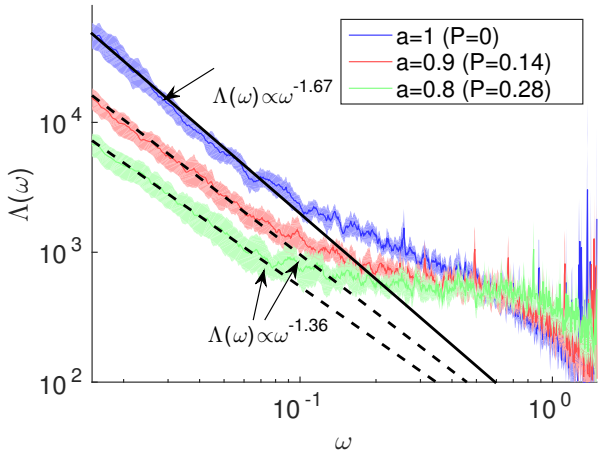}
 \caption{Mean free path $\Lambda(\omega)$ versus frequency for different pressures $P$ at mean temperature $T=0.2$. The MFPs are plotted in units of $a$, i.e., reflecting the decay length as a function of particle number. The temperature bias in the nonequilibrium simulation was $\Delta T=0.1$.}
 \label{fig:Lambdas_p}
 \end{center}
\end{figure}

Figure \ref{fig:Lambdas_p} shows the MFPs at different pressures for the mean temperature $T=0.2$. Exerting pressure on the chain can be seen to suppress the MFPs at low frequencies. Fitting a power-law scaling $\Lambda(\omega)\propto \omega^{-\eta}$ to MFPs at low frequencies suggests that the divergence exponent $\eta$ is reduced by the pressure as predicted by non-linear hydrodynamics. The fitting exponents at zero ($\eta\approx 1.67$) and non-zero ($\eta \approx 1.36$) pressure differ, however, from the values $\eta=(1-\alpha)^{-1}$ predicted by non-linear fluctuational hydrodynamics ($\eta=2$ and $\eta=1.5$, respectively). Because it is difficult to resolve the exact values of the exponents from our numerical simulations, we cannot rule out the possibility that the disagreement between our simulations and nonlinear hydrodynamics only arises from numerical inaccuracies, but the observed differences both for $P=0$ and $P\neq 0$ could also suggest the breakdown of non-linear fluctuational hydrodynamics.

Figure \ref{fig:Lambdas_p} also shows that the pressure reduces the frequency-dependence of the MFPs, generating a flatter $\Lambda$ vs. $\omega$ curve.  Similar flattening of $\Lambda(\omega)$ has been observed also in the relaxation times calculated for three-dimensional silicon \cite{parrish15}. 

\section{Conclusions}
\label{sec:conclusions}

With the emergence of low-dimensional systems and their potential applications in thermal devices, understanding anomalous thermal conduction remains an important goal in thermal physics. In this paper, we have determined the frequency-dependent phonon mean free paths (MFPs) from nonequilibrium simulations in the simplest system exhibiting anomalous conduction, an anharmonic $\beta$-FPU chain with quartic non-linearity. In contrast to other methods, the determined MFPs directly reflect the resistance to heat flow at each vibrational frequency, allowing for directly estimating the contributions of different vibrational frequencies to thermal conduction in chains of different lengths. 

The determined MFPs diverge in a power-law fashion $\Lambda(\omega) \propto \omega^{-\eta}$ at low frequencies, in harmony with diverging thermal conductivity. Linear fitting delivers the value $\eta=1.67$, in agreement with mode-coupling theory \cite{lepri98} and thermal conductivity divergence $\kappa(N)\propto N^{0.40}$. It is, however, challenging to fully establish the statistical error in the exponent $\eta$, and longer simulations with larger system sizes are needed in future to resolve the exact value. At very low frequencies, the determined nonequilibrium MFPs agree with the equilibrium MFPs determined in an earlier work \cite{liu14}, but the two MFPs differ at higher frequencies. Exerting pressure on the chain reduces the MFPs at low frequencies, with a reduction also in the value of $\eta$. The determined exponents differ from the values expected from non-linear fluctuational hydrodynamics \cite{mendl13,spohn14}, but the reasons for this disagreement remain unclear. Simulating longer chains and determining nonequilibrium MFPs also for different systems is expected to deliver much needed insight into thermal conduction in low-dimensional systems and to even solve the remaining disputes regarding anomalous conduction. For a more transparent picture of the microscopic origins of the low-frequency MFP divergence, full solution of the Boltzmann transport equation with microscopic phonon-phonon scattering rates could prove useful.  

\section{Acknowledgements}
We thank Stefano Lepri for discussions. The computational resources were provided by the Finnish IT Center for Science and the Aalto Science-IT project. The work was partially funded by the Emil Aaltonen foundation, Aalto Energy Efficiency Research Programme (AEF) and the Academy of Finland.

\appendix

\section{Effects of the bath parameters}
\label{app:A}

 \begin{figure}[bt]
  \begin{center}
   \includegraphics[width=8.6cm]{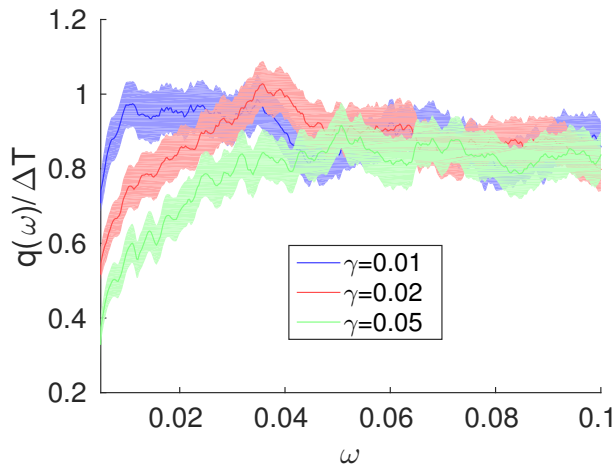}
  \caption{Normalized spectral current for $N=1000$, $N_b=1000$, $T=0.2$, $\Delta T=0.1$ with various values of the bath coupling constant $\gamma$.}
  \label{fig:app1}
  \end{center}
 \end{figure}

This appendix briefly discusses the effects of the bath parameters $N_b$ and $\gamma$ on the results. When choosing the bath relaxation constant $\gamma$, the value should be small enough to avoid (i) generating acoustic mismatch between the leads and the center region \cite{saaskilahti13} and (ii) violently disturbing phonon dynamics in the leads by too strong dissipation. To investigate the dependence of spectral heat currents on $\gamma$, we show in Fig. \ref{fig:app1} the spectral current at low frequencies for $N=1000$ with various values of $\gamma$. The results of Fig. \ref{fig:app1} show that the spectral current is independent of $\gamma$ at all except very low frequencies. For frequencies $\omega\lesssim \gamma$, current starts to decrease, most likely due to the overdamping of modes. In the results of Sec. \ref{sec:results}, where we use $\gamma=0.01$, we do not therefore show results for $\omega < \gamma = 0.01$. Choosing a smaller value of $\gamma$ would allow for probing smaller frequencies, but this would also require increasing the size $N_b$ of the thermalized regions.

 \begin{figure}
  \begin{center}
  \includegraphics[width=8.6cm]{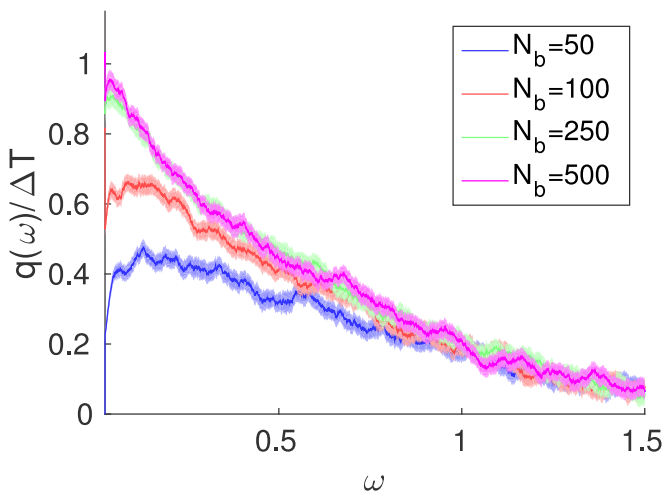}
  \caption{Normalized spectral current for $N=1000$, $\gamma=0.01$, $T=0.2$, $\Delta T=0.1$ with various values of the bath length $N_b$. Spectral smoothing has been carried out with spectral width $\Delta \omega=0.05$.}
  \label{fig:app2}
  \end{center}
 \end{figure}

When choosing the length $N_b$ of the thermalized regions, the goal is to minimize the contact resistance between the baths and the chain so that the heat current is only limited by the \textit{internal} conductance of the chain. Therefore, $N_b$ should be chosen as large as possible, effectively mimicking the semi-infinite lead configuration of the Landauer-B\"uttiker setup (see, e.g., Ref. \cite{saaskilahti13}). To illustrate the dependence of the spectral current on $N_b$, we plot in Fig. \ref{fig:app2} the spectral current for $N=1000$ with various values of $N_b$. The results show that choosing $N_b\gtrsim 3\gamma^{-1}$ is sufficient to eliminate the finite-size effects at low frequencies. In the results of Sec. \ref{sec:results}, we choose $N_b=1000$.

%


\end{document}